\documentclass[twocolumn]{aastex63}
\usepackage{CJK}

\usepackage{graphics}

\def\msun{\,  {\rm M_\odot}}
\def\cm-3{\,{\rm cm^{-3}}}

\def\kpc-3{\,{\rm kpc^{-3}}}
\def\myr-1{\,{\rm Myr^{-1}}}

\def\kpc{\,{\rm kpc}}

\def\tc{$t_{\rm{c,0}}$}
\def\tcg{$t_{\rm{c,1}}$}
\def\rfade{$R_{\rm{fade}}$}
\def\tmulti{$t_{\rm{multi}}$}

\def\fpri{$f_{\rm{unh}}$}

\def\fcolor{$f_{\rm{color}}$}

\usepackage{natbib}
\usepackage{color}
\usepackage{rotating}
\usepackage{footnote}
\usepackage{datetime}
\usepackage{amsmath}
\usepackage{mathrsfs}
\usepackage[normalem]{ulem}
\submitjournal{ApJ}

\shorttitle{SNe Ia Feedback \& Cool Phase Formation}
\shortauthors{Li et al.}

\graphicspath{{./}{figures/}}

\begin{document}
\begin{CJK*}{UTF8}{gbsn}

\title{The Impact of Type Ia Supernovae in Quiescent Galaxies: \\
I. Formation of the Multiphase Interstellar medium}

\correspondingauthor{Miao Li }
\email{mli@flatironinstitute.org}

\author[0000-0003-0773-582X]{Miao Li   
(李邈)}
\affiliation{Center for Computational Astrophysics, Flatiron Institute, New York, NY 10010, USA}

\author[0000-0001-5262-6150]{Yuan Li (黎原)}
\affiliation{Center for Computational Astrophysics, Flatiron Institute, New York, NY 10010, USA}
\affiliation{Department of Astronomy, and Theoretical Astrophysics Center, University of California, Berkeley, CA 94720, USA}

\author[0000-0003-2630-9228]{Greg L. Bryan}
\affiliation{Center for Computational Astrophysics, Flatiron Institute, New York, NY 10010, USA}
\affiliation{Department of Astronomy, Columbia University, 550 West 120th Street, New York, NY 10027, USA}

\author[0000-0002-0509-9113]{Eve C. Ostriker}
\affiliation{Department of Astrophysical Sciences, Princeton University, Princeton, NJ 08544, USA}

\author[0000-0001-9185-5044]{Eliot Quataert} 
\affiliation{Department of Astronomy, and Theoretical Astrophysics Center, University of California, Berkeley, CA 94720, USA}

\begin{abstract}
A cool phase of the interstellar medium has been observed in many giant elliptical galaxies, but its origin remains unclear. We propose that uneven heating from Type Ia supernovae (SNe Ia), together with radiative cooling, can lead to the formation of the cool phase. The basic idea is that since SNe Ia explode randomly, gas parcels which are not directly heated by SN shocks will cool, forming multiphase gas. We run a series of idealized high-resolution numerical simulations, and find that cool gas develops even when the overall SNe heating rate $H$ exceeds the cooling rate $C$ by a factor as large as 1.4. We also find that the time for multiphase gas development depends on the gas temperature. When the medium has a temperature $T = 3\times 10^6$ K, the cool phase forms within one cooling time \tc; however, the cool phase formation is delayed to a few times \tc\ for higher temperatures. The main reason for the delay is turbulent mixing. Cool gas formed this way would naturally have a metallicity lower than that of the hot medium. For constant $H/C$, there is more turbulent mixing for higher temperature gas. We note that this mechanism of producing cool gas cannot be captured in cosmological simulations, which usually fail to resolve individual SN remnants. 

\end{abstract}

\keywords{Type Ia supernovae (1728); Core-collapse supernovae (304); Elliptical
galaxies (456); Galaxy evolution (594); Interstellar medium (847); Hydrodynamical simulations (767); Shocks
(2086); Galaxy chemical evolution (580) }

\section{Introduction}
\label{intro} 
Most massive elliptical galaxies in today's universe are thought to be ``red and dead'': most stars are old and the ongoing star formation activity is very low. The interstellar medium (ISM) is mainly hot and tenuous, with $T> 10^6$ K. However, cool\footnote{In this paper we refer to  $T \lesssim 10^4$ K as ``cool" gas} gas and dust have been increasingly detected in a significant fraction of elliptical galaxies \citep{phillips86,vandokkum95,knapp96,macchetto96,sarzi06,combes07,young11,alatalo13,davis16}. 
Volume-limited surveys of elliptical galaxies find that 20-30\% of the systems have CO emission, with estimated molecular masses in the range $10^7 -10^9 M_\odot$ \citep{welch10, young11}. An ionized phase with temperature around $10^4$ K is also detected, although the mass is much less \citep{kim89,buson93,goudfrooij94,macchetto96,pandya17}.

The origin of the cool gas is unclear. Several scenarios have been proposed. Externally, galaxy mergers and filamentary accretion may bring in fresh cool gas \citep{sarzi06,sarzi10,young11,pandya17}. Internally, mass loss from evolved stars may supply cool gas \citep{parriott08,li19}. Cold gas may also form out of the hot phase due to local thermal instabilities \citep{Balbus1989}. Observationally, the morphology of the hot gas is more irregular in systems hosting cool ISM, with a lower entropy at the outskirts of elliptical galaxies, suggesting that the thermal instability is at work \citep[e.g.][]{werner14}.

There have been a lot of theoretical/numerical studies on thermal instabilities in hot halos in massive galaxies and galaxy clusters \citep[see][and references therein]{Voit17}. Since most of the massive systems are observed to be in rough global thermal equilibrium (with no classical cooling flow), most of the numerical simulations either superimpose global equilibrium \citep{McCourt2012, Sharma2012} or achieve a quasi-steady state where AGN heating globally balances cooling \citep{Gaspari2012, yli15, wang19}. In these simulations, thermal instabilities may develop because of a local imbalance between cooling and heating \citep{Meece2015}, AGN jet uplifting \citep{Li2014}, and turbulence \citep{voit18}.

Feedback from Type Ia Supernovae (SNe Ia) explosions is important in keeping the gas hot in elliptical galaxies. It has been shown that the total energy from SNe Ia is on the same order of magnitude as the radiative cooling \cite{voit15}. SNe Ia are from the old stellar populations, and their locations are random. Each explosion releases roughly $10^{51}$ ergs of energy, which drives a blast wave into the ISM, heating the gas. Many SNe Ia together maintain a hot ISM, and may even drive global outflows \citep{tang09}.

In a series of two papers, we study the feedback from SNe Ia on the hot ISM of elliptical galaxies. In this paper, we propose that the random heating by SNe Ia, together with radiative cooling, would result in multiphase gas, even when the overall heating rate $H$ is moderately larger than the cooling rate $C$. 
This mechanism can only be captured when individual SN remnants are resolved ($\sim$10 pc), which is beyond the capability of current cosmological simulations (resolutions $\gtrsim$ kpc). Thus these simulations can miss an important mechanism to produce cool gas in the ISM. In a companion paper (Paper II), we investigate the energetics and turbulence of the medium under the impact of SNe Ia. We find that SNe Ia provide unique feedback to the hot ISM, which is much more complex than a volumetric heating source, as is commonly assumed. The majority of their impacts are not shown when SNe are under-resolved.

We organize our paper as follows. In Section 2, we estimate analytically the amount of gas that will cool down, due to the uneven heating from SNe. In Section 3 we describe the setup of a set of numerical experiments to test the proposed idea. We then present our simulation results in Section 4. We discuss the implications for galaxy evolution and cosmological simulations in Section 5, and conclude in Section 6. 

This is a paper from the \textit{Simulating Multiscale Astrophysics to Understand Galaxies} (SMAUG) collaboration\footnote{\url{www.simonsfoundation.org/flatiron/center-for-computational-astrophysics/galaxy-formation/smaug}}, a project intended to improve models of galaxy formation and large-scale structure by working to understand the small-scale physical processes that cannot yet be directly modeled in cosmological simulations.

\section{Formation of cool phase due to uneven heating of Supernovae}
\label{sec:analytic}

In this Section, we propose the basic picture for the formation of cool phase due to the uneven heating of Type Ia SNe. We give a quantitative estimate of the time when cool gas forms and how much cool gas there will be. We then show that this can happen even when the overall SNe heating rate $H$ is larger than the radiative cooling rate $C$.

The basic idea is that each SN Ia only heats a finite volume of gas. Since SNe Ia are the end results of old stellar evolution, the locations of the explosions follow the old stellar population, but are otherwise located randomly. Because of the stochastic nature, some part of the hot ISM may not be heated by any SNe within its cooling time, thus it will cool (assuming other heating processes are negligible). 

Before we proceed to the quantitative calculation, we first emphasize that the general evolution of SNR in a hot, tenuous medium is quite different from that in the warm/cool ISM, typical of spiral galaxies. Unlike the later case, a SNR in a hot medium does not produce a cooling shell. This excludes the possibility that the cool gas is formed near the shock front of each individual SNR. The standard picture of a SN-driven spherical shock includes four evolutionary phases: (i) free expansion, when the SN ejecta moves ballistically; (ii) Sedov-Taylor phase, during which the total energy is conserved; (iii) cooling stage, when radiative cooling becomes important and a thin shell forms at shock front; (iv) fadeaway, when the blast wave fades into a sound wave \citep{chevalier74}. In a medium that is cool and relatively dense, like the ISM of a spiral galaxy, all four stages are present. 
In a hot and tenuous medium, however, the cooling is usually inefficient and the sound speed of the medium is large, so the blast wave decays into a sound wave before cooling becomes significant. As a result, the formation of a thin shell is missing \citep{mathews90,tang05}.  
In star-forming spiral galaxies where the ISM temperature is $10^4$ K or lower, a cool shell forms when the shock velocity drops to about 100 km $s^{-1}$, and when the post-shock temperature is close to the peak of the cooling curve. So when the sound speed of the ISM is larger than 100 km $s^{-1}$, the pressure equilibrium is reached before cool shell formation.

Derived in detail in Appendix \ref{sec:shell}, we find that the ISM condition for the absence of a shell formation is that the density of the ISM is less than a critical value, which is temperature-dependent,
\begin{equation}
n< n_{\rm{crit}}= 0.11\ \rm{cm^{-3}}\  (T_6/\alpha)^{3.85} E_{\rm{51}}^{-0.50},
\end{equation}
where $T_6 = T/10^6$ K, $E_{\rm{51}} = E_{\rm{SN}}/ 10^{51} \rm{erg} $, and $\alpha$ is a free parameter of order unity. This condition is generally satisfied for the hot ISM of elliptical galaxies, some galactic bulges of spiral galaxies, and galaxy clusters.

The blast wave driven by a SNR fades into a sound wave when the pressure inside the SN bubble is comparable to that of the ambient medium, that is,  
\begin{equation}
\alpha (\gamma -1 ) \frac{E_{\rm{SN}}}{V_{\rm{SN}}} = P,
\end{equation}
where $P$ is the pressure of the ambient gas, $E_{\rm{SN}}$ is the energy released by a SN, $V_{\rm{SN}}$ is the volume occupied by the blast wave, and $\gamma$ is the adiabatic index. For a fully ionized plasma, $\gamma = 5/3$. The corresponding radius when fade-away happens in a uniform medium with a number density $n$ and temperature $T$ is then 
\begin{equation}
\begin{split}
R_{\rm{fade}} & =  \left( \alpha (\gamma-1) \frac{E_{\rm{SN}}}{4\pi P/3} \right) ^{1/3} \\
& = 48.8 \rm{pc}\ \alpha ^{1/3} \ E_{51} ^ {1/3} n_{0.02} ^{-1/3} T_7 ^{-1/3}, 
\end{split}
\label{eq:rfade}
\end{equation}
where $T_7 = T/10^7$K, $n_{0.02} = n/0.02 $ cm$^{-3}$, and $E_{\rm{51}} = E_{\rm{SN}}/10^{51}$erg.

The medium has a cooling rate $C=n^2 \Lambda$, where $\Lambda n$ is the cooling rate per proton. Assuming collision ionization equilibrium, $\Lambda$ only depends on temperature. The isochoric cooling time of the medium is
\begin{equation}
t_{\rm{c,0}} (n, T) = \frac{P}{(\gamma-1) n^2 \Lambda (T)}. 
\label{eq:tcool}
\end{equation}

Assuming SNe explode at a constant rate $S$ (per volume per time), the volume-averaged heating rate is $H=SE_{\rm{SN}}$. Since the locations of SNe are independent, the fraction of unheated gas, $f_{\rm{unh}}$, averaged over many realizations, decreases with time \citep{mckee77},
\begin{equation}
    df_{\rm{unh}}/dt=-f_{\rm{unh}} SV_{\rm{SN}}.
    \label{eq:df}
\end{equation}
Solving this equation gives an exponential decline of $f_{\rm{unh}}$ (assuming $t=0$ is when SNe start to explode):
\begin{equation}
f_{\rm{unh}} = e^ {-t SV_{\rm{SN}}} = e^ { - \frac{t}{t_{\rm{c,0}}}\frac{H}{C}   }.
\label{eq:f_unheated}
\end{equation}
This indicates that, after $t=$\tc, $e^{-H/C}$ of the volume will not be covered by any SN bubbles. Note that we assume here that blast waves are the dominant heating source. Sound waves can in principle heat the gas, too \citep{fabian17}, but we neglect this effect in this paper. As a result, the unheated gas will cool down, and the medium becomes multiphase. Furthermore, the cool phase can form even when $H/C>1$. Note that $H$ and $C$ are spatially-averaged quantities over a volume $\gg V_{\rm{SN}}$. Also, $H$, $C$, \tc\ here are for the initial uniform condition. As the system evolves, the medium will become inhomogeneous and these values will not be the same everywhere. However, the inhomogeneity will only occur once a large fraction of the volume is covered by SNe. For simplicity, we assume that they are constant as a first step.

\begin{figure}
\begin{center}
\includegraphics[width=0.50\textwidth]{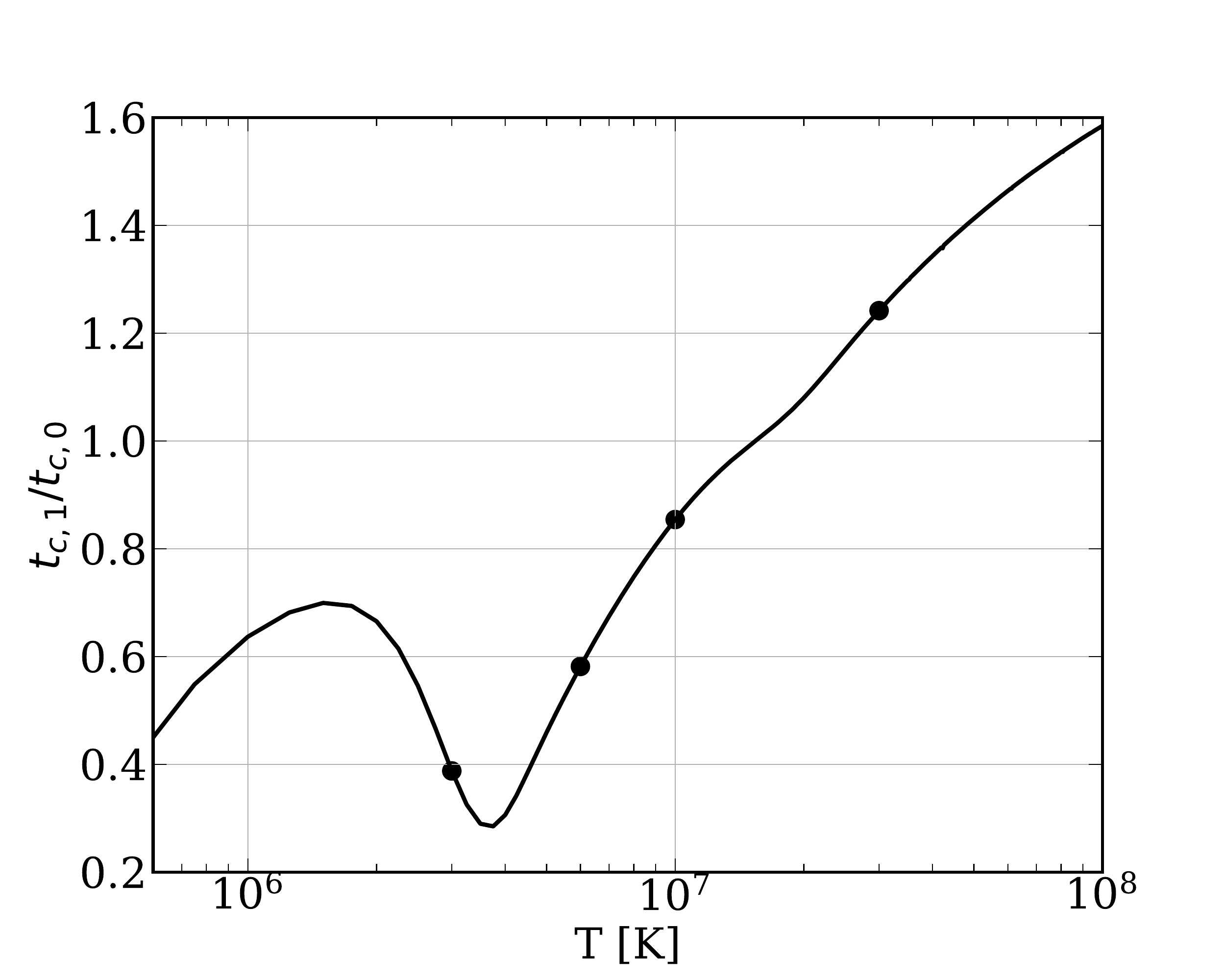}
\caption{The ratio between two cooling time as a function of temperature. The time for gas to cool to $2\times 10^4$ K, $t_{\rm{c,1}}$, is smaller than the instantaneous cooling time, \tc, when $T < 1.3 \times 10^7$ K.  The four dots show the initial temperatures of the medium in our numerical experiments. }
\label{f:t_cool_ratio}
\end{center}
\end{figure}

When a parcel of gas is not heated, the time for it to form a cool phase ($T_c \lesssim 10^4$ K) is not exactly \tc. This is because during the cooling process, the gas temperature decreases, so the cooling rate changes as well. More precisely, we should use the cooling time that is the integral of the cooling rate over the range of $T$ as gas cools, that is,  
\begin{equation}
t_{\rm{c,1}} (n, T) = \int_{Tc}^{T} \frac{3k_b dT'}{2n \Lambda(T')}, 
\label{eq:tc1}
\end{equation}
The coefficient $3/2$ assumes the cooling is isochoric; in the isobaric limit, the coefficient should be replaced by $5/2$. For the cooling curve we adopt in this paper, the ratio of $t_{\rm{c,1}}/t_{\rm{c,0}}$ is shown in Fig. \ref{f:t_cool_ratio}. 
In the calculation, we take $T_c = 2\times 10^4$ K. The ratio is not sensitive to the exact value of $T_c$, as long as it is around $10^4$ K. 

The integrated cooling time \tcg\ determines when the cool phase starts to occur. For a cool phase formation cycle, the mass fraction of cool phase as a function of time is a step function, 
\begin{equation}
     f_c =
    \begin{cases}
      0, & \text{for}\ t< t_{c,1} \\
      e^ { - \frac{t_{\rm{c,1}} }{t_{\rm{c,0}}}\frac{H}{C}   } , &\text{for}\ t \geqslant t_{c,1}.
    \end{cases}
    \label{eq:fc}
\end{equation}
As soon as the cool phase develops, the remaining hot gas expands to fill the space, and the above cycle starts again (with different $C$, \tc\, and $t_{\rm{c,1}}$ since the mean density of hot gas is lower).
Eq. \ref{eq:fc} indicates that if $t_{\rm{c,1}}/t_{\rm{c,0}} < 1$, the formation of the cool phase is further promoted by having an earlier onset than \tc\ and a larger fraction of cool gas than $e^{-H/C}$. Conversely, if $t_{\rm{c,1}}/t_{\rm{c,0}} > 1$, the development of cool phase is delayed later than \tc\ and the amount of cool gas is less than $e^{-H/C}$.

The above calculations are based on simplified assumptions; in particular, we assume that the medium is uniform as in the initial conditions and remains static. Of course these are not true in a realistic ISM, and in particular, SNe themselves can change the underlying gas properties. In the next section, we use hydrodynamic simulations to check the validity of the proposed mechanism and examine quantitatively the evolution of ISM under SNe Ia feedback.

\section{Numerical Setup}
\label{sec:method}

\linespread{0.9}
\begin{table}[]
\begin{center}
\caption{Model Parameters}
\label{table1}
\setlength\tabcolsep{0.5pt}
\begin{tabular}{cccccccc}
\hline
Name             & S                       & d     & $R_{\rm{fade}}$ & $t_{\rm{c,0}}$ & $t_{\rm{c,1}}$ & $t_{\rm{multi}}$ & $t_d$ \\

                 & {\footnotesize (Mpc$^{-1}$ } & (kpc) & (pc)            & (Myr)          & (Myr)          & (Myr)            & (Myr)    \\
&{\footnotesize $\cdot$ kpc$^{-3}$ )} & & & & & &    \\
                 \hline
n0.32-T1e7-H0.8C & 776                     & 0.3   & 19.4            & 33             & 28             & 72               & 7.0    \\
n0.32-T1e7       & 990                     & 0.3   & 19.4            & 33             & 28             & 85               & 6.0                  \\
n0.32-T1e7-H1.2C & 1165                    & 0.3   & 19.4            & 33             & 28             & 85               & 5.3                  \\
n0.32-T1e7-H1.4C & 1359                    & 0.3   & 19.4            & 33             & 28             & --               & 4.7                 \\\hline
n0.16-T1e7       & 248                     & 0.5   & 24.5            & 67             & 57             & 180              & 11.0                \\
n0.08-T1e7       & 61.9                    & 1     & 30.8            & 133            & 114            & 375              & 19.0                 \\
n0.02-T1e7       & 3.87                    & 4    & 48.8            & 534            & 456            & 1700             & 61.0                 \\\hline
n0.08-T3e6       & 110                     & 1     & 46.0            & 22             & 9              & 18               & 6.0                 \\
n0.08-T6e6       & 51.6                    & 1     & 36.5            & 96             & 56             & 180              & 16.5                \\
n0.08-T3e7       & 92.3                    & 1     & 21.3            & 265            & 329            & 1162             & 25.0                 \\\hline
n0.02-T3e6       & 6.85                    & 4    & 72.9            & 90             & 35             & 79               & 20.0                 \\
n0.02-T3e6-hr    & 6.85                    & 4    & 72.9            & 90             & 35             & 82               & 28.0                 \\
n0.02-T3e6-H1.1C & 7.39                    & 4    & 72.9            & 90             & 35             & 84               & 19.0                 \\
n0.02-T3e6-H1.2C & 8.06                    & 4    & 72.9            & 90             & 35             & 95               & 18.0                 \\
n0.02-T3e6-H1.4C & 9.40                    & 4    & 72.9            & 90             & 35             & 155              & 16.0                 \\
n0.02-T3e6-H1.8C & 12.1                    & 4    & 72.9            & 90             & 35             & --               & 13.0                 \\\hline
n0.002-T3e6      & 0.0685                 & 20    & 157.2           & 905            & 351            & 850              & 130.0      \\\hline

\end{tabular}
\end{center}
\end{table}

Now we turn to numerical simulations with full hydrodynamics to test the mechanism of forming cool gas proposed in Section 2. 

The simulations are performed using the Eulerian hydrodynamical code \textsc{Enzo} \citep{bryan14}. The boundary conditions are periodic for all three directions. We use the finite-volume piece-wise parabolic method \citep{colella84} as the hydro-solver. For the radiative cooling, we use the cooling curve from \cite{rosen95}, for the temperature range of $300-10^9$ K, assuming a gas metallicity of a half solar value. The cooling curve is shown in Appendix A (Fig. \ref{f:cooling_curve}). The resolution and box size scale with $R_{\rm{fade}}$ for the initial medium.
Each SN is implemented as injecting $E_{\rm{SN}}=10^{51}$ erg thermal energy, $m_{\rm{SN}}=$ 1 $\msun$, and a fixed amount of ``color'', $m_{\rm{color}}$, a tracer fluid (passive scalar) that follows the mass. These quantities are evenly distributed in a sphere, with an injection radius 0.5$R_{\rm{fade}}$. The metal ejection of SNe Ia would change the cooling rate, but in this work we did not include the metal-dependent cooling for simplicity. 
The resolution is chosen so that $R_{\rm{fade}}$ at the beginning of the simulation is equal to the length of six computational cells. Since $R_{\rm{fade}}$ is resolved in our simulations, the blast wave will automatically evolve into the Sedov-Taylor solution, with an energy partition of roughly 30\% kinetic and 70\% thermal. Each box size is 20 $R_{\rm{fade}}$.

We start each simulation with a uniform, static medium, with a gas number density $n$, and a temperature, $T$. We list all our simulations in Table 1.  The input parameters of $n$ and $T$ are based on the observed range of hot gas in elliptical galaxies. The third column of Table \ref{table1} lists the typical galactocentric distance for the corresponding $n$ in observed giant elliptical galaxies \citep{voit15}.
SNe are injected at random locations in the box, with a constant frequency $S$. The value of $S$ for each run is listed in Table 1. In reality, $S$ depends on the local stellar density (and age of those stars) \cite{cappellaro99,pain02,maoz17}. For a rough estimate, $S$ scales with the stellar density $\rho_*$ as $S \sim 5\times 10^3$ Myr kpc$^{-3}$ ($\rho_*/10^{11}$ M$_\odot$ kpc$^{-3}$) \citep{scannapieco05}.
The overall SNe Ia heating rate in the box is thus $H\equiv S E_{\rm{SN}}$. This is in comparison with the radiative cooling rate of initial condition, $C\equiv n^2 \Lambda (T)$. 
There is evidence for $H/C \sim$ 1: Observationally, thermal balance is seen in giant elliptical galaxies for a wide range of radii \citep[e.g.][]{voit15}. Theoretically, if $H/C$ is out of the equilibrium, the system will adjust itself. For example, if $H>C$, gas will be heated and expand, where the extra heat is converted into motions; if $H<C$, gas will cool and the density of hot gas decreases until $H\sim C$ in the hot phase \citep[e.g.][]{maller04}.   
For the fiducial runs, we set $S$ so that $H/C=1.02$. We also vary $H/C$, to see how results depend on this parameter.

Each run is represented by a name that encodes its $n$, $T$, and $H/C$. For example, ``n0.02-T1e7-H0.8C'' means $n=0.02 $ cm$^{-3}$, $T=10^7$K, and $H=0.8C$. If $H/C$ is not indicated, then the fiducial value 1.02 is used.  The discretization of the computational cells affects the energy of each SN at the 1\% level. Each simulation runs for 4\tc\ (\tc\ is also calculated for the initial $n$ and $T$).  

We emphasize that we use these simplified and idealized setups as a proof of concept to explore the new mechanism. We will discuss the implications of the missing physics and future improvements in Section~\ref{sec:discussion}.

\section{Results}

We have carried out a set of simulations with different $n$ and $T$ in order to find whether multiphase gas appears and when. Before examining the general trend systematically, it is worth looking at two representative cases to understand the basic evolution of the gas.

\subsection{Case 1: $T = 3\times 10^6$ K}

\begin{figure}
\begin{center}
\includegraphics[width=0.5\textwidth]{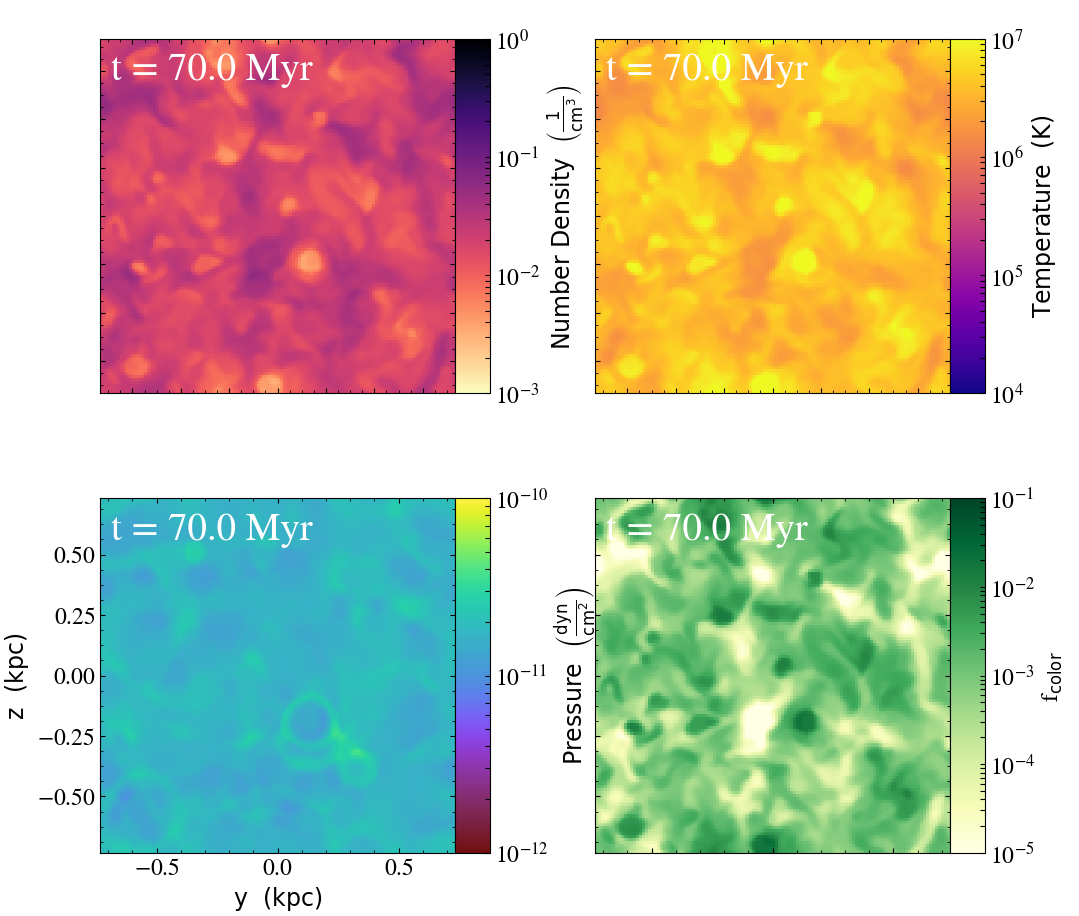}
\includegraphics[width=0.5\textwidth]{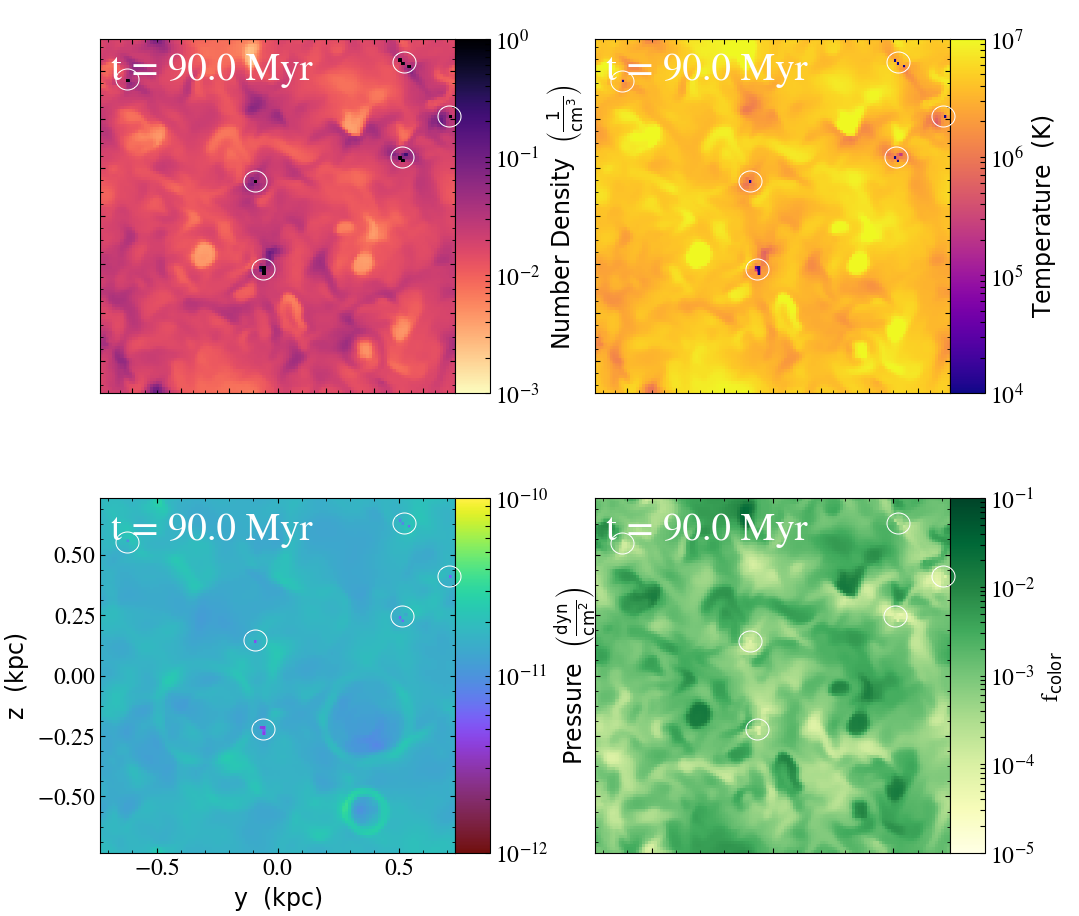}
\caption{ Slices of number density, color, pressure, and \fcolor\ for n0.02-T3e6, right before ($t=70$ Myr, upper four panels) and after ($t=90$ Myr, lower four panels) the multiphase formation. Cool gas (centered on by the white circles) forms preferentially in regions with relatively high densities, low temperatures and low \fcolor. }
\label{f:slice_3e6K}
\end{center}
\end{figure}

We use the run n0.02-T3e6 as an example to illustrate the $3\times 10^6$ K case. The cooling time for the medium at the beginning of the simulation is \tc$=$ 90 Myr. We define the time when any gas parcel cools below $2\times 10^4$ K as \tmulti, which is 75 Myr for this run.

The upper four panels of Fig. \ref{f:slice_3e6K} show slices of density, temperature, pressure and color fraction at 70 Myr, right before the formation of cool gas. The SN color fraction, \fcolor, is defined as the ratio between the color density and gas density (similar to metallicity), normalized to that of the SN ejecta. The medium is visually inhomogeneous, with signs of SN bubbles. The spatial variation of density and temperature is about a factor of 10. The pressure varies much less than density or temperature, although spherical blast waves/sound waves driven by SNe are visible. It is most obvious in the \fcolor\ panel which part of gas has been mixed with SNe ejecta: areas with darker shades have been polluted by SN ``colors'', whereas regions with light shades have not been impacted by SNe directly. Comparing \fcolor\ with the other panels, one can see that low \fcolor\ corresponds to higher densities and lower temperatures, and vice versa. 

The lower panel of Fig. \ref{f:slice_3e6K} shows slices at $t=90$ Myr, after the cool phase forms. The cool gas is seen as tiny clumps that have the highest densities and lowest temperatures (centered on white circles). 
Comparing to the plots at $t=70$ Myr, one can see that the cool phase forms in regions with low \fcolor. Note that the cool clumps do not form at the shock front of individual SNR.

\begin{figure}
\begin{center}
\includegraphics[width=0.50\textwidth]{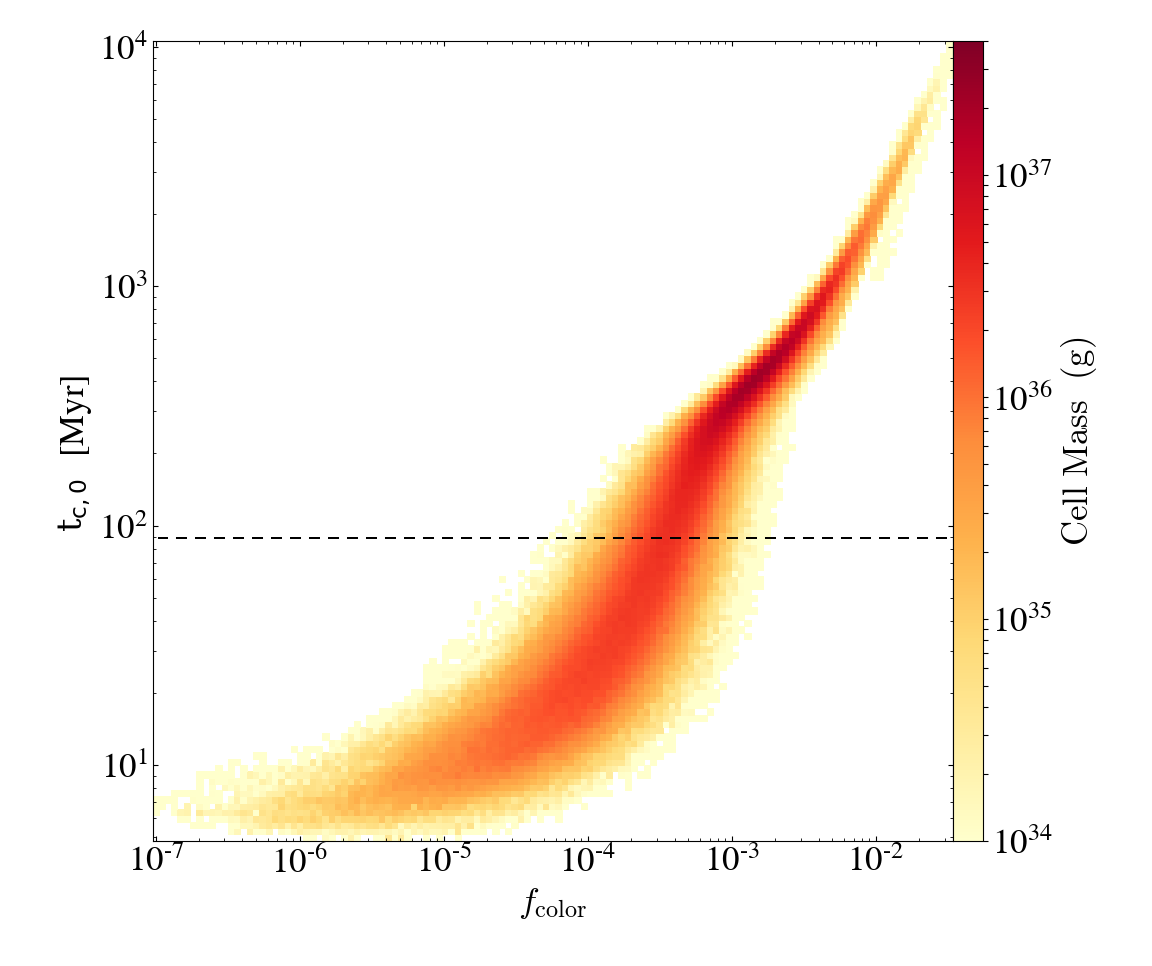}
\includegraphics[width=0.50\textwidth]{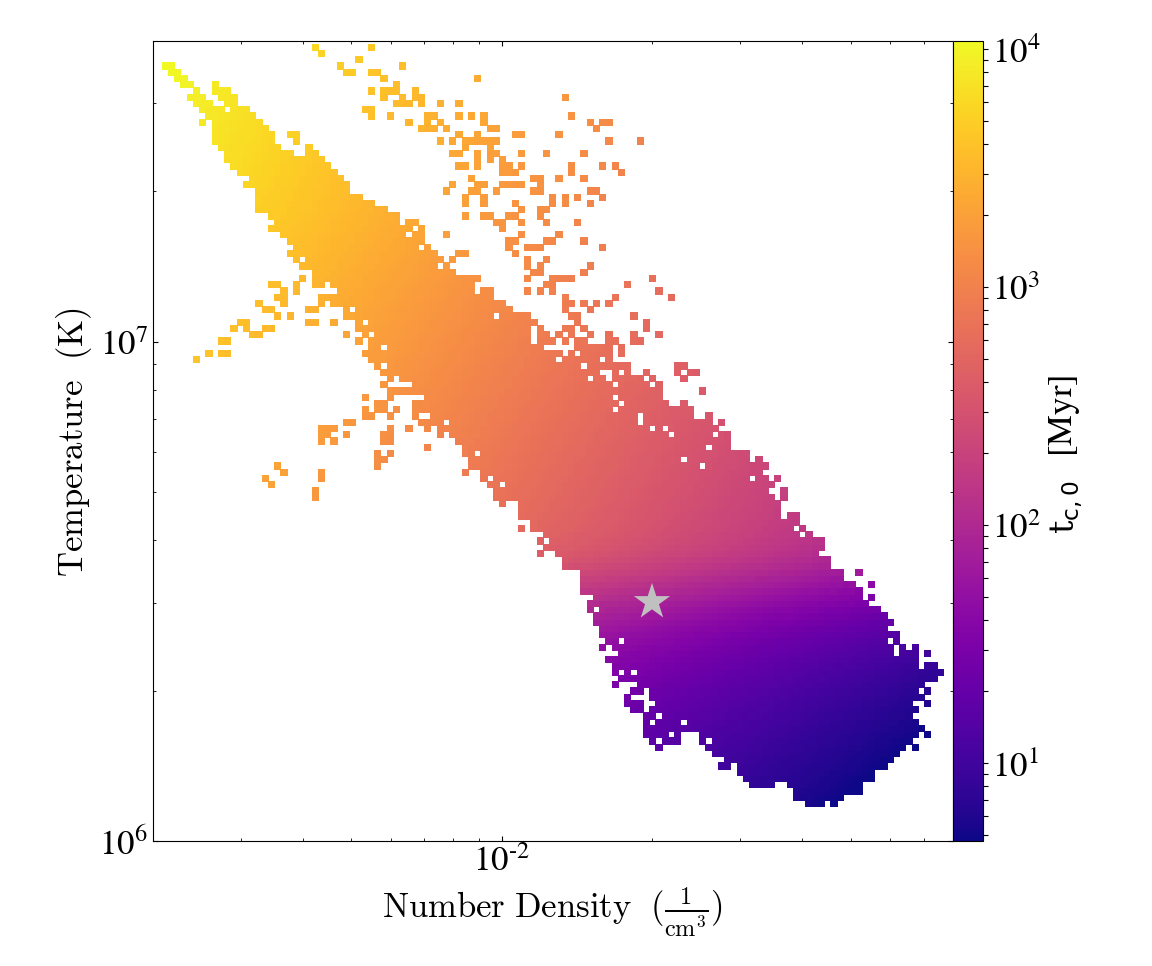}
\caption{Phase diagram for the run n0.02-T3e6K at $t=$70 Myr, right before the formation of the cool phase. The dotted line in the upper panel indicates \tc\ for gas at $t=0$.  The star in the lower panel indicates the initial condition of the gas. }
\label{f:color_tcool_3e6K}
\end{center}
\end{figure}

To better see the correlation of physical properties of the inhomogeneous gas, we plotted in Fig. \ref{f:color_tcool_3e6K} two phase diagrams at $t=$70 Myr, before the cool phase formation (same time as the first snapshot of Fig.~\ref{f:slice_3e6K}).
The upper panel shows the isochoric cooling time \tc\ versus \fcolor, color-coded by the total mass. At the beginning of the simulation, all gas resides in a single point on the diagram: \tc $=$ 90 Myr and \fcolor $=$0. The dashed line shows the initial \tc. At $t=$70 Myr, \tc\ spans about 3 orders of magnitude, and \fcolor\ spans 5 orders of magnitude. There is a fairly tight correlation between  \tc\ and \fcolor: gas with long \tc\ tends to have higher \fcolor. This is easy to understand: the heating and rarefaction effect of the blast waves make hot, tenuous bubbles, which arrive at a rough pressure balance with the ambient medium at \rfade. For $T\gtrsim 10^5$ K, at fixed pressure, \tc is shorter for lower temperatures.
New SNe bubbles have the highest \fcolor\ and longest \tc. The tightest correlation of \tc\ and \fcolor\, seen in the upper-right part of the plot, arises from these new SNe bubbles. 
The majority of the gas has \tc $>$ 90 Myr, but some gas has \tc $\ll$ 90 Myr. The short-\tc\ tail will form the cool phase within 10 Myr. 

The lower panel of Fig. \ref{f:color_tcool_3e6K} shows density versus temperature, color-coded by the isochoric \tc. The star symbol indicates the initial condition. 
The distribution of temperature and density lie roughly along a constant pressure. Gas now has temperatures and densities both above and below the original value. Gas with the shorter cooling time (darker color shade) has higher densities and lower temperatures. 
These features are consistent with the expectations described in Section 2. The medium becomes inhomogeneous under the impact of Type Ia SNe. The gas that is going to become part of the cool phase (i.e. has the shortest \tc) has the lowest \fcolor, largest density, and lowest temperature. This gas has been outside of any SNe bubbles. Also note that the overall gas pressure is somewhat higher than the initial condition, so the cooling process is not entirely isobaric. This can cause the cooling time to slightly deviate from \tcg.

\begin{figure}
\begin{center}
\includegraphics[width=0.50\textwidth]{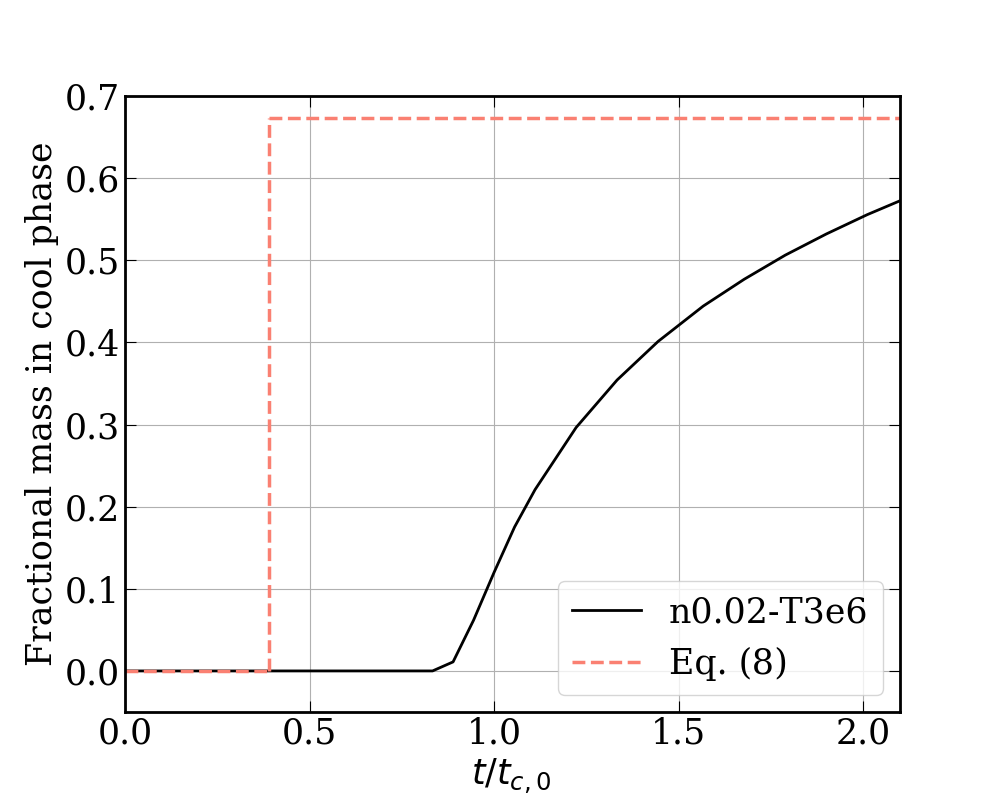}
\caption{ Fractional mass in the cool phase ($T <2\times 10^4$ K) as a function of time for the run n0.02-T3e6. The time is normalized by \tc. The dashed line is the step function in Eq. \ref{eq:fc} for one cycle of developing cool phase. The mismatch of the two curves indicates that the gas cools more slowly than the simple expectation where all gas unheated by SNe cools at \tcg. }
\label{f:fmcool_002_3e6K}
\end{center}
\end{figure}

Now we turn to more quantitative analysis, by examining when cool gas starts to form and how much cool gas forms with time. 
Fig. \ref{f:fmcool_002_3e6K} shows the fraction of mass in the box that is in the cool phase ($T <2\times 10^4$ K), as a function of time. The cool phase occurs at about 0.8\tc. After that, the mass of the cool phase appears to accumulate rapidly: by \tc, 13\% of mass in the box is cool; by 2\tc, about 55\% of the mass is cool. We warn, however, that the accumulation rate of the cool phase may not be accurate, since (1) the cool phase, once formed, is not well resolved, (2) some relevant physics -- such as thermal conduction and sound-wave heating (sound waves can steepen into shocks after entering the cool gas) -- are not included or captured, and (3) we have neglected the larger context and gravitational field of the elliptical. 
The dashed line indicates the theoretical estimate from Eq. \ref{eq:fc}. Plugging in relevant parameters, Eq. \ref{eq:fc} shows that the onset of cool gas formation is predicted at 0.30\tc, and 67\% of the mass would be in cool phase after that. The simulation results, on the other hand, show a later start of multiphase gas formation, and a slower accumulation of cool mass over time. This suggests that the simple expectation where all gas unheated by SNe cools at \tcg has something missing, and possibly that hydrodynamical effects are playing a role. We will discuss this further in Section~\ref{sec:discussion}.

\subsection{Case 2:  $T=10^7$ K} 

\begin{figure}
\begin{center}
\includegraphics[width=0.50\textwidth]{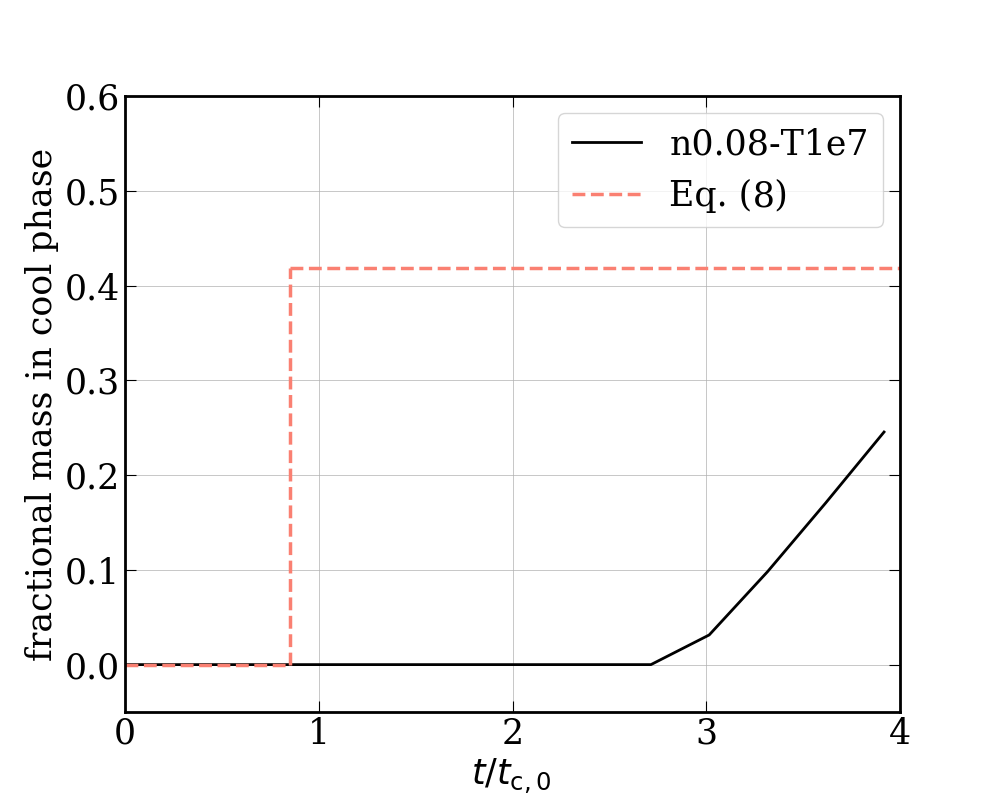}
\caption{Same as Fig. \ref{f:fmcool_002_3e6K}, but for n0.08-T1e7. }
\label{f:fmcool_t_008_1e7K}
\end{center}
\end{figure}

Now we discuss a run with a higher initial temperature, $10^7$ K. We find that the multiphase gas is, once again, produced, but not until 2-3\tc has elapsed, in contrast to the lower temperature case, where the cool phase occurs before \tc.  We use the run $\rm{n0.08-T1e7}$ as an example. Such gas has an isochoric \tc$=$133 Myr. 
Fig. \ref{f:fmcool_t_008_1e7K} shows the fractional mass in the cool phase as a function of time (similar to Fig. \ref{f:fmcool_002_3e6K}). As one can see, the cool phase does not form until $t\approx 2.7$\tc, which is about $375$ Myr. By the end of the simulation, only 27\% of mass is in the cool phase. 
It is expected, from Eq. \ref{eq:fc}, that the cool phase formation time would be later, and the mass fraction of cool gas is less, compared to the lower-temperature run of n0.02-T3e6. For n0.08-T1e7, \tcg$\approx$ 0.85\tc. Therefore, according to Eq. \ref{eq:fc}, the cool phase would form at 0.85\tc with a mass fraction of 0.42, comparing to 0.39\tc\ and a mass fraction of 0.67 for n0.02-T3e6. Quantitatively, however, the actually formation time of cool gas in the simulation is much more delayed.

\begin{figure}
\begin{center}
\includegraphics[width=0.50\textwidth]{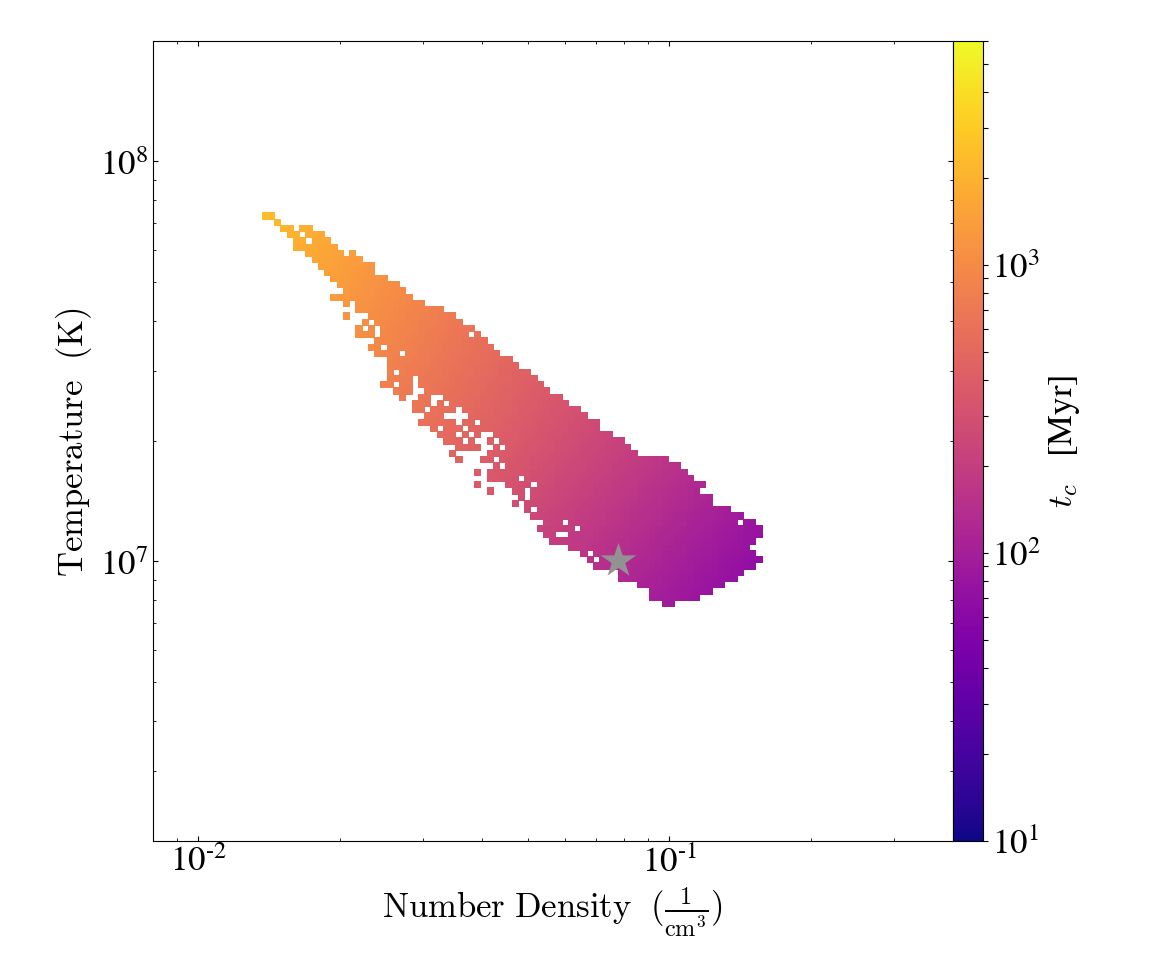}
\includegraphics[width=0.50\textwidth]{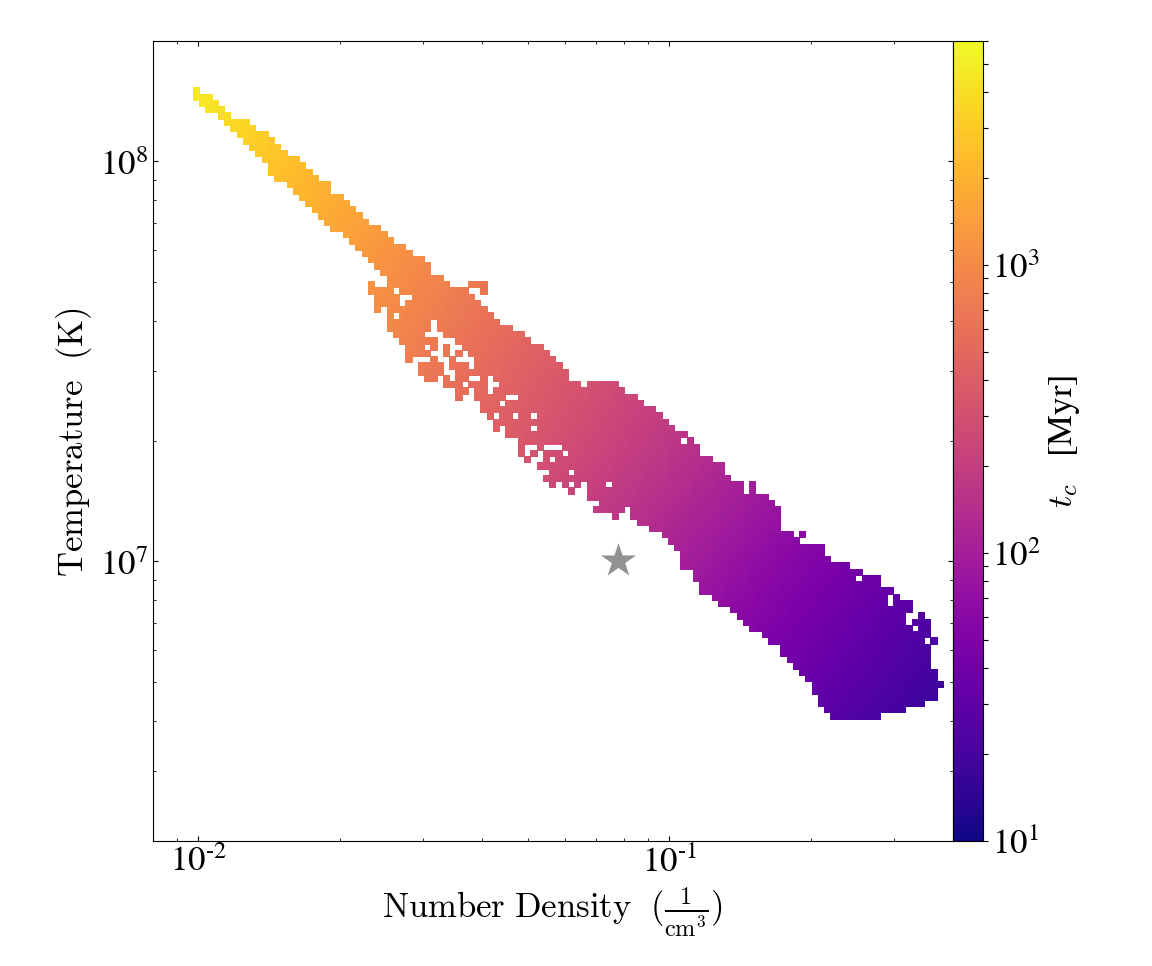}
\caption{Same as the lower panel of Fig. \ref{f:color_tcool_3e6K}, but for the run n0.08-T1e7. The upper panel is for $t=120$ Myr (slightly before \tc$=$133 Myr), and the lower is for $t=340$ Myr (somewhat before \tmulti$=$370 Myr). The stars indicate the initial condition. }
\label{f:n_T_008_1e7K}
\end{center}
\end{figure}

To better understand what is causing the delay, we plotted in Fig. \ref{f:n_T_008_1e7K} the phase diagram of temperature versus density, color-coded by the isochoric \tc\ at two times in this simulation. The upper panel is for $t=$120 Myr, slightly before \tc (133 Myr), and the lower panel is for $t=$340 Myr, somewhat before the cool phase forms (375 Myr). The star symbols indicate the initial condition of the gas. At 120 Myr, the distribution of gas density and temperature do broaden over time, similar to what was seen for the n0.02-T3e6 run. That said, the shortest \tc\, which is around 90 Myr, does not fall much below the initial value of 133 Myr. This means that not much gas has lost sufficient thermal energy to form a cool phase soon. The majority of the gas has a higher temperature and lower density, which corresponds to a longer \tc. Only at a much later time (lower panel) does some gas reaches a short enough \tc. At 340 Myr, the distribution of density and temperature lie above the initial condition. This implies that the overall gas has been heated considerably, and thus the thermal pressure is above the initial value.

\subsection{A parameter survey for multiphase gas}

\begin{figure}
\begin{center}
\includegraphics[width=0.5\textwidth]{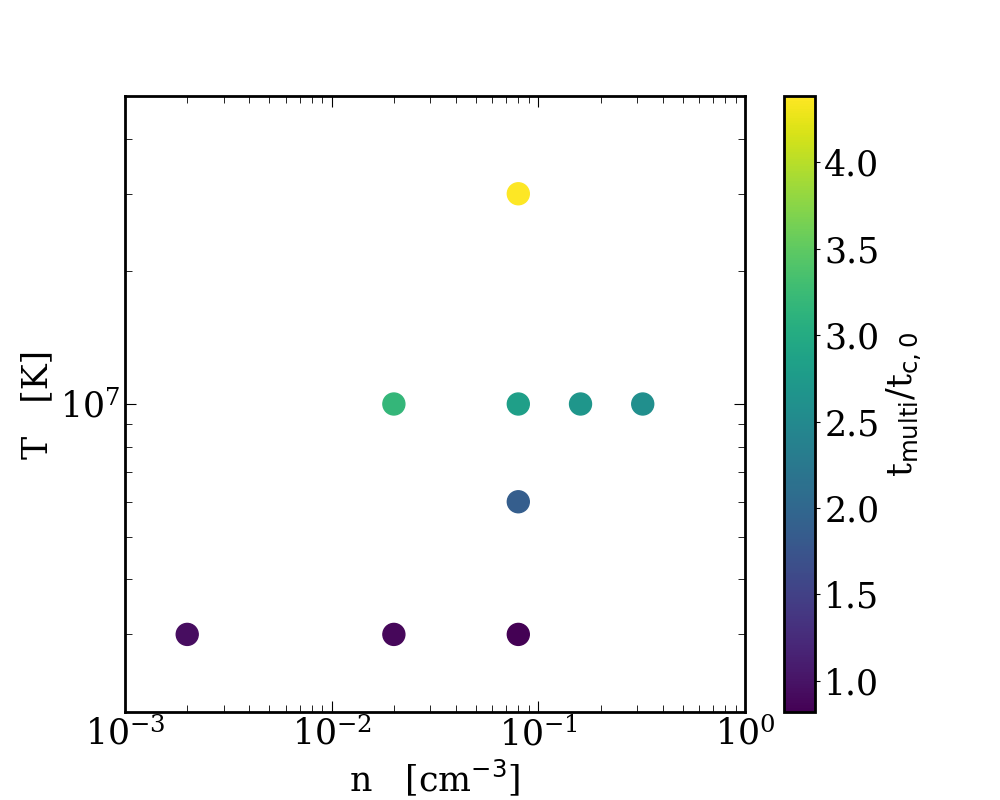}
\caption{ All simulations with H/C=1.02. Each dot represents one simulation. The x- and y-axis show the density and temperature of the ambient gas. The color shows the ratio between the formation time of cool phase, \tmulti, and the instantaneous cooling time, \tc. The ratio increases with temperature of the ambient gas. }
\label{f:Ia_all_t_multiphase_t_cool}
\end{center}
\end{figure}

Besides the illustrated cases with $T=3\times 10^6$K and $10^7$ K above, we extend the parameter space to a wide range of $n$, and, in particular, have two additional simulations with  $T=6\times 10^6\ $K and $3\times 10^7\ $K with the same $n=$ 0.02 cm$^{-3}$. All these simulations have the fiducial $H/C=$1.02. We find that all these runs eventually develop multiphase gas, demonstrating the universality of such an outcome, even under the uneven heating of Type Ia SNe. 

The time at which the cool gas forms, though, varies. Fig. \ref{f:Ia_all_t_multiphase_t_cool} shows \tmulti$/$\tc\ for all the simulations. The ratio \tmulti$/$\tc\ increases monotonically with increasing $T$. For the case with the highest temperature $T=3\times 10^7$ K, cool gas forms at about 4.4\tc. For the same $T$, the dependence of \tmulti$/$\tc\ on $n$ is quite small. For example, all runs with $T=3\times 10^6$ K develop a cool phase within \tc, even when $n$ varies by a factor of 40. This is also shown by the four runs with $T=10^7$ K: when $n$ increases by a factor of 16, \tmulti$/$\tc\ only decreases very mildly from 3.2 to 2.6. 
We will discuss the possible reason for the delay of cooling in the next section.

\begin{figure}
\begin{center}
\includegraphics[width=0.5\textwidth]{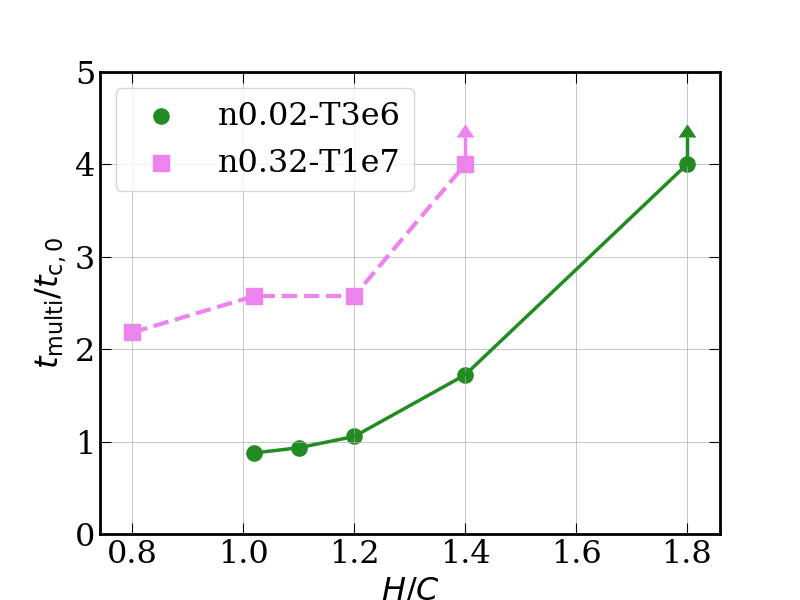}
\caption{Formation time of multiphase \tmulti\ for different vales of $H/C$. The lower limit arrows indicate no cool gas formed within 4\tc. The development of cool phase is further delayed for larger $H/C$.
}
\label{f:t_multi}
\end{center}
\end{figure}

The fiducial run assumes a rough balance of heating and cooling, i.e., $H/C=$1.02. How robust is the formation of cool phase to an increase in $H/C$? To investigate that, we vary the SN rate for n0.02-T3e6 and n0.32-T1e7. 
We vary $H/C$ from 0.8-1.8. 
Fig. \ref{f:t_multi} shows the time when the multiphase gas forms, \tmulti\, as a function of $H/C$. For n0.02-T3e6, the multiphase is persistent when H/C is $\leqslant$ 1.4. The time when the cool phase first occurs is delayed as $H/C$ increases. When $H/C= $1.4, \tmulti\ is almost twice the value for $H/C=$1.02.  When H/C$=$ 1.8, the cool phase does not occur within the duration of simulation, which is 4 \tc. For n0.32-T1e7, a similar trend with $H/C$ is seen. When $H/C=1.4$, \tmulti$>$ 4\tc. Note that even when $H/C=0.8$, i.e., the cooling rate is higher than the heating, the cool phase still does not develop until after 2\tc. 

The correlation between \tc$/$\tmulti/ and $H/C$ suggests that forming cool phase is more difficult with more intense heating, which makes intuitive sense. Though, from Eq. \ref{eq:fc}, the onset of the cool phase should not depend on $H/C$. We will discuss the delay of cooling in more detail in Section \ref{sec:reason}.

\subsection{Turbulent mixing}
\label{sec:reason}

We found in the previous section that the formation time of the cool phase is delayed from the idealized calculation of Eq. \ref{eq:fc}. The delay is longer when the medium has a higher temperature. In this section, we will discuss turbulent mixing as an important cause of this delay. 

Feedback from SNe not only heats the gas, but also induces motions. In a static, uniform medium, SN explosions will drive spherical, outward shocks. Due to the pressure of the ambient medium, the blast waves will decay into sound waves. As the blast and/or sound waves from explosions at different locations interact with each other, the medium becomes turbulent. Turbulence tends to mix material and energy, therefore transporting heat from SN bubbles to the ambient medium. Gas that is not directly heated by SN-driven blast waves can be heated by this turbulent heat transport. With this extra heating, it takes longer for the gas to cool down. Indeed, turbulent mixing has been suggested as a way to suppress thermal instability \citep[e.g.][]{banerjee14}. In our cases, the cool phase does form eventually, but we argue that mixing postpones the onset of it.

\begin{figure}
\begin{center}
\includegraphics[width=0.50\textwidth]{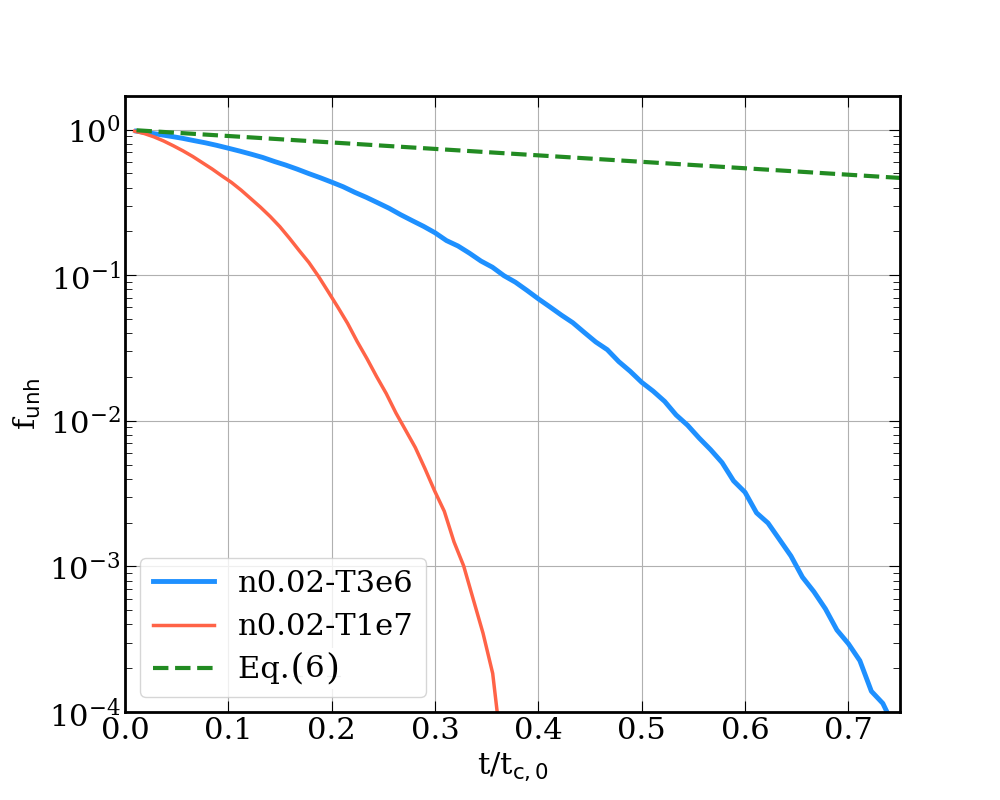}
\caption{Volume fraction of gas unpolluted by SNe, \fpri, as a function of time.}
\label{f:color_fraction}
\end{center}
\end{figure}

To quantify turbulent mixing, we use ``color'' to trace where SN ejecta goes, which is a proxy for the region that is covered by SN bubbles\footnote{Strictly speaking, ``color'' does not trace SN heating precisely. ``Color'' tracks the SNe ejecta, which is confined by the contact discontinuity, whereas the energy of SNe is passed on to all shocked regions. Since the shock front runs ahead of the contact discontinuity, the shocked volume is larger than the ``color''-polluted one. In reality, though, the blast wave pushes the majority of shocked material into a thin layer behind the shock front \cite{sedov59}, therefore the contact discontinuity may not be well-resolved. The shock front and the ``color'' front are almost identical in practice. When the shock wave decays into a sound wave, the ``color'' front stops moving, while the sound wave keeps going outward; therefore, the two fronts decouple.}. If there is no mixing at all, and the domain of SN heating is approximated by a sphere with a radius \rfade, then according to Eq. \ref{eq:f_unheated}, the volume fraction that is not heated by any SN bubble, \fpri, decays exponentially with time.

 With turbulent motions, the actual \fpri declines faster than what is predicted by Eq. \ref{eq:f_unheated}. To find gas that is not heated by any SN bubbles, we have chosen a sufficiently small value of SN color density as a cutoff, 
\begin{equation}
  \rho_{\rm{color}} = \kappa m_{\rm{color}}/V_{\rm{SN}} = \kappa m_{\rm{color}} P/E_{\rm{SN}}, 
\end{equation}
where $\kappa = 4.4\times 10^{-4}$. This chosen value of $\kappa$ is sufficiently small that the results are insensitive to the precise value used. In Fig. \ref{f:color_fraction}, the dashed line shows \fpri\ as a function of time according to Eq. \ref{eq:f_unheated}, while the solid lines show the actual \fpri, for the two example runs, n0.02-T3e6, and n0.02-T3e7. The solid lines are well below the dashed line. Furthermore, the way that \fpri\ declines deviates from a simple power-law and declines much faster at later time. We use the timescale $t_d$ to empirically quantify the decline of \fpri, which we take to be the time when \fpri\ equals $1/e$. 

For n0.02-T3e6, $t_d=$ 20 Myr, and \tc$/t_d \approx$ 4.5; for n0.02-T1e7, $t_d =$ 61 Myr, and \tc$/t_d \approx$ 9. The ratio \tc$/t_d$ indicates the effectiveness of mixing as a way to spread SN heating, relative to radiative cooling. Larger ratios indicate more efficient mixing, leading to a longer delay in the development of the cool phase. 

\begin{figure}
\begin{center}
\includegraphics[width=0.5\textwidth]{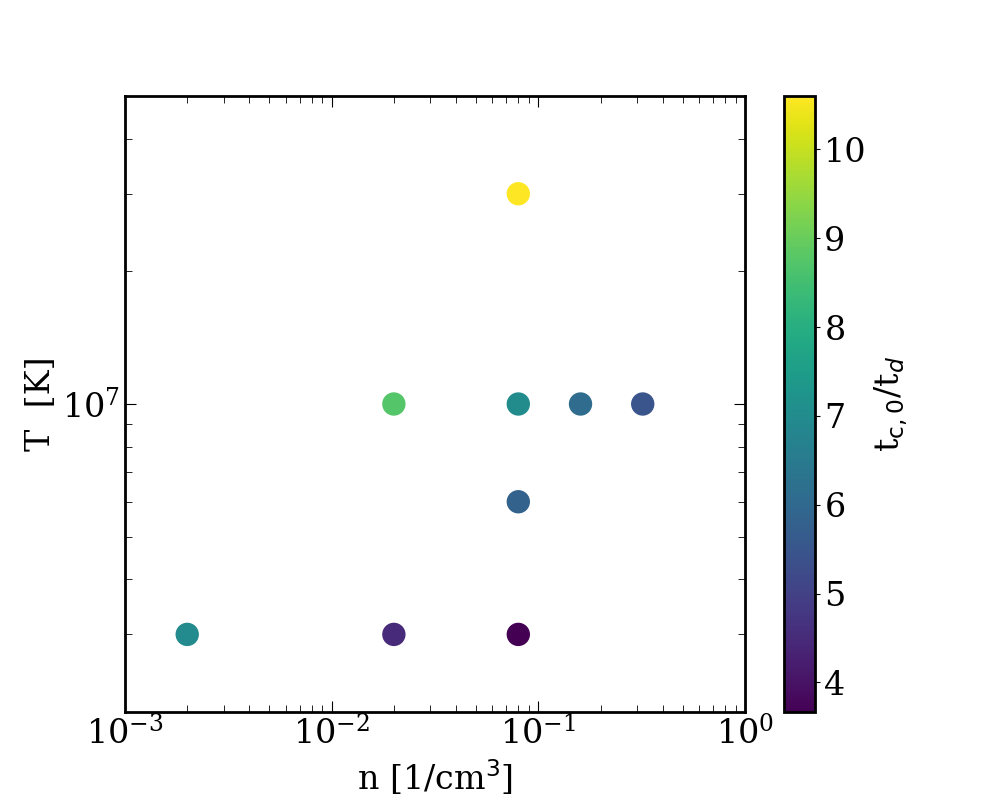}
\caption{Ratio of the cooling time to the empirical mixing time, \tc$/t_d$ for all simulations with $H/C=1.02$. Each simulation is represented by one circle on the scatter plot. The two axes indicate the two input parameters of each run. The ratio increases with temperature, suggesting that mixing is stronger for hotter medium. }
\label{f:tc_td}
\end{center}
\end{figure}

Fig. \ref{f:tc_td} shows \tc$/t_d$ for all simulations with $H/C=1.02$. 
The value of \tc$/t_d$ ranges from 4-12, indicating that $t_d$ is much smaller than \tc\ and that turbulent mixing is generally important. 
The ratio increases with increasing temperature and decreasing density, but is more sensitive to the former. For example, when increasing the temperature by a factor of 10 for the runs with n0.08, from 3$\times 10^6$ to 3$\times 10^7$ K, the ratio changes from 3.7 to 10.6. When increasing the density by a factor of 16 for the run with $T=10^7$ K, the ratio only decreases mildly from 8.8 to 5.5. This trend is very similar to what is found in Fig. \ref{f:Ia_all_t_multiphase_t_cool}. 
The similarity of the patterns in Fig. \ref{f:Ia_all_t_multiphase_t_cool} and \ref{f:tc_td} indicates that turbulent mixing is effective in delaying formation of cool phase for a higher temperature medium. The increase of \tc$/t_d$ also is also seen when $H/C$ is larger, confirming that mixing contributes to the delayed cooling shown in Fig. \ref{f:t_multi}.

In this paper, we have established the relation between the delayed cooling and the turbulent mixing by measuring $t_d$ empirically. In Paper II, where we investigate the turbulence structure of hot ISM in detail, we will present a theoretical model for $t_d$ based on simple physical arguments.

\section{discussion}
\label{sec:discussion}

In this paper, we propose a new mechanism of forming cool gas in a hot ISM of quiescent galaxies and galaxy clusters. The cool gas can form due to the stochastic heating of SNe Ia. 
Through numerical simulations, we demonstrate that a multiphase gas forms universally, at least under our idealized setups, although turbulent mixing is playing a role in delaying the cooling to a few times the cooling time. 
In this section, we discuss what is still missing from our idealized simulations, the ramifications of our results to the evolution of the systems and to observations, and implications for cosmological simulations.

\subsection{Missing physics}

As discussed in Section \ref{sec:reason}, turbulent mixing can transport heat to gas that is not directly heated by SN bubbles, therefore smoothing the uneven heating by random SNe. In our idealized simulation setup, the only driver of turbulence is SNe themselves. In real galaxies, other processes can also contribute to the turbulence, such as galaxy mergers \citep{paul09,iapichino17} and AGN feedback \citep{gaspari12b,valentini15,wang19}. These processes may contribute significantly to mixing depending on the magnitude of the turbulence they drive. Unfortunately, turbulent velocities and the relevant driving scales are poorly constrained by current observations of the hot gas in elliptical galaxies \citep{ogorzalek17}.
In Paper II, we will discuss the SNe-driven turbulence in these systems in more detail, and compare them to the theoretical estimates of the turbulence driven by other processes. 

Thermal conduction is another mechanism of heat transport \citep{spitzer56}. Conduction may thus delay the formation of the cool phase. After the formation of the cool phase, depending on the size of cool clumps, thermal conduction can result in evaporation of the clouds or condensation of hot gas onto the cloud \citep{cowie77}. However, our resolution is far coarser than the Field length, therefore thermal conduction cannot be self-consistently modeled. Additionally, the cool clumps are as small as a few cells, thus not well-resolved in these simulations. So the amount of cool gas once multiphase gas appears may not be accurate. Dedicated simulations resolving the conductive layer and cool clumps are needed to further study this problem \citep[e.g.][]{borkowski90,ferrara93,bruggen16,liang20}.

\begin{figure}
\begin{center}
\includegraphics[width=0.50\textwidth]{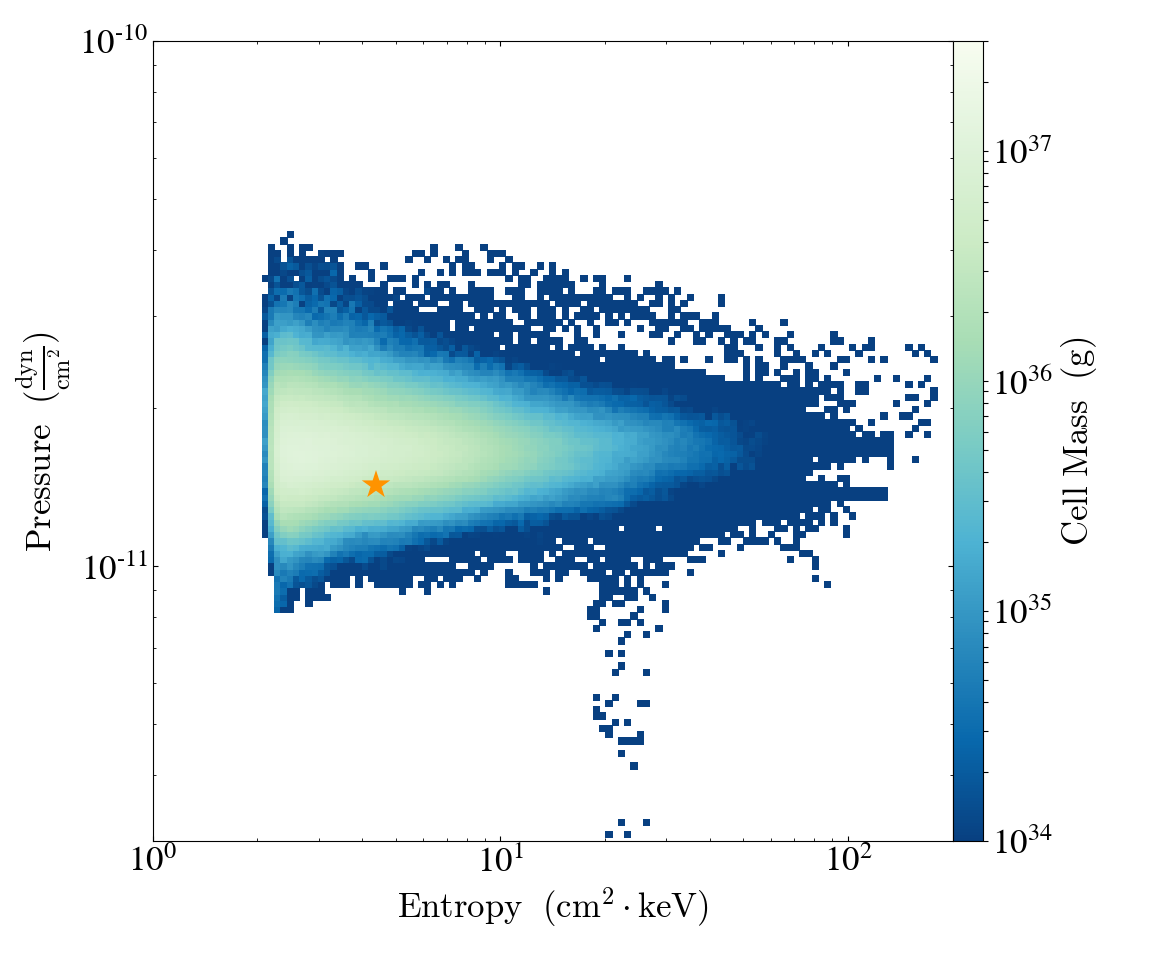}
\caption{Pressure versus specific entropy at $t=$50 Myr ($\approx$ 0.56\tc) for n0.02-T3e6. The star indicates the initial condition. The pressure is largely in equilibrium while the entropy shows a large span by  more than one order of magnitude. The inhomogeneity in entropy implies that gas may stratify in presence of gravity. }
\label{f:entropy_pressure}
\end{center}
\end{figure}

Our simulations only model a patch of the ISM in elliptical galaxies. Gravitational fields and gas stratification are not included. 
In a realistic environment with gravitational stratification, SN bubbles have a lower density than the ambient medium and tend to rise because of the buoyancy force; similarly, the unheated gas has a relatively higher density and tends to sink \citep{mathews90}. Fig. \ref{f:entropy_pressure} shows the phase diagram of gas pressure and specific entropy for n0.02-T3e6. The snapshot is taken at t$=$50 Myr, which is about 0.56 \tc, when the cool phase has not formed yet. The initial condition is indicated by the star. The pressure has a very narrow range within a factor of a few. But the specific entropy already spans a factor of $>$ 30. The range is similar to the observed range for radii over two logarithmic scales: $R=0.1-10$ kpc \citep[see Fig. 1 of][]{voit15}. When stratification is present, a gas parcel that has a different specific entropy from the environment will move in the gravitational potential, until its entropy matches that of the ambient medium (if not considering mixing and cooling/heating).

The effect of buoyancy damping on thermal instability in a stratified medium has been studied extensively in recent years \citep[see][and references therein]{Voit17}. Due to the lack of gravity in our simulations, the conditions that we find for thermal instability to develop can be considered necessary but not sufficient in real galaxies with gravity. That is, if buoyancy damping is effective (i.e., when the cooling timescale is significantly longer than the dynamical timescale), the instability that can grow in our idealized box will likely be damped via buoyancy oscillation in a stratified medium.

One implication of the stratification is the metal transport. These high-entropy SN bubbles are metal-enriched, and if mixing is not efficient, the rising of SN bubbles may preferentially distribute metals outside the central regions of the galaxy \citep{tang09,voit15}, and even to the circumgalactic medium \citep{zahedy19}. In addition, by the time we stopped the simulations, the multiphase ISM is still evolving and does not reach a steady state. Therefore, the quantitative properties of the simulated ISM may not be representative of real systems. This also poses an obstacle for comparing the simulations directly to the observations. These effects cannot be captured in our simulations with periodic boundary conditions, but would be very interesting to investigate in future studies that include gravity and gas stratification.

\subsection{Ramifications of the results}

A cool phase can form due to the random heating of SNe Ia. The cool clumps, once formed, will fall toward the center of the galaxy/galaxy cluster. Cool gas produced by this mechanism may be an important source for the observed multiphase systems. The cool gas could then form stars, and/or be accreted to the central supermassive black hole and fuel the active galactic nuclei \citep{ciotti07,choi12,yli15,yuan18}. These processes consume the cool phase, and in turn inject energy back to the system.

Since SNe Ia heat and metal-enrich the medium at the same time, the chemical abundance of the cool phase would be smaller than that of the hot medium, since the cool phase develops as a result of being missed by any SNe. The mean abundance difference between gas covered by SN bubbles and the cool phase is $\Delta Z = M_{\rm{met,Ia}}/M(R_{\rm{fade}})$, the ratio between the metal mass ejected by a SN Ia and the enclosed mass within \rfade. Assuming the cool phase has a metallicity (metal mass fraction) of $Z_0$ and using Eq. \ref{eq:rfade} for \rfade, the percentage difference of metallicity is thus
\begin{equation}
\begin{split}
    \frac{\Delta Z}{Z_0} & = \frac{(\gamma -1) E_{\rm{SN}} M_{\rm{met,Ia}}}{c_s^2 Z_0} \\
   & = 0.27 \left(\frac{T}{10^7 K}\right) \left(\frac{0.01}{Z_0}\right) \left(\frac{M_{\rm{met,Ia}}}{1 M_\odot}\right),
    \end{split}
    \label{eq:delta_Z}
\end{equation}
where $E_{\rm{SN}}=10^{51}$ erg is assumed for the second equality. This equation can apply to either the metallicity including all elements, or a certain element (such as iron) when $Z$ and $M_{\rm{met,Ia}}$ are replaced by the corresponding quantities for that element. 
Eq. \ref{eq:delta_Z} emphasizes the abundance difference between the hot and the cool, and indicates that the difference is more prominent when the ISM is hotter and/or has a smaller $Z_0$. In addition, the cool phase forms at different epochs and locations of galaxies would have different chemical abundances. This at least partly explains the observed diversity of metallicity of cool gas in early-type galaxies \citep[e.g.][]{sarzi06}.

\subsection{Implications for cosmological simulations}

In current cosmological simulations, due to the coarse resolution, SNe feedback can only be included as a subgrid model. As a result, the above mechanism for forming a cool phase due to uneven heating is not captured. The modeling of feedback from SNe Ia in these simulations is quite simple, usually as a heating term. The amount of heating for each volume (or mass) element in one time step is just the total energy from SNe that would have exploded within that time step, which is related to the stellar density and age. If the added SNe energy is larger than the energy loss due to radiative cooling (n$^2\Lambda\Delta t$), then the thermal energy of this element will increase, and will not form any cool gas.  
Therefore, cosmological simulations miss an important way to produce the cool phase. Furthermore, due to the large inhomogeneity of the hot gas (Fig. \ref{f:color_tcool_3e6K} and \ref{f:n_T_008_1e7K}), the actual cooling rate of the medium deviates from $n^2\Lambda (T)$, where $n$ and $T$ are the averaged quantities (averaged over the large length and math scales resolved in such simulations).  This effect is investigated in more detail in Paper II. 

To capture the real impact of SNe feedback, the resolution requirement is that \rfade be resolved by at least a few computational cells. Note that for a typical warm/cool ISM in spiral galaxies, one needs to resolve the cooling radius $R_{\rm{cool}}$ to capture SN feedback \citep{simpson15,kim15,li15}. But in a hot ISM in elliptical galaxies, \rfade $\ll R_{\rm{cool}}$ (Appendix 3), thus a general condition for the resolution would be resolving min(\rfade,$R_{\rm{cool}}$) by 3-10 cells. 
For typical conditions in elliptical galaxies, \rfade\ ranges from 20-150 pc (Table~\ref{table1}), meaning that the spatial resolution has to be a fraction of that, which is beyond current computing capabilities. That said, doing so in smaller systems like galactic bulges and dwarf elliptical systems are possible \citep[e.g.][]{tang09}; and thanks to the rapid development of computing facilities, simulating the whole massive elliptical system with resolved SNe Ia feedback would be possible within a few years.

\section{Conclusions}

In this paper, we propose that uneven heating by randomly-located Type Ia SNe is a mechanism that can produce a cool phase in the hot ISM of elliptical galaxies. We run a series of idealized numerical simulations modeling a part of the hot ISM to test this idea. The major conclusions are the following: 

(1) With radiative cooling, uneven heating from Type Ia SNe almost always leads to the formation of a cool phase. Gas that is not heated by SN bubbles will cool down in roughly the cooling timescale \tc. This mechanism is at work even when the overall heating rate, $H$, is larger than the radiative cooling rate, $C$.

(2) When the hot medium has a higher temperature, the formation time of the cool phase is delayed until a few cooling times have elapsed.

(3) The formation of the cool phase is delayed as $H/C$ increases. Yet the cool phase still develops within 4\tc\ when $H$ is as high as 1.2-1.4$C$.

(4) Turbulent mixing transports heat from SN bubbles to the ambient medium. We attribute the delay of the cooling to an increased level of turbulent mixing. 

(5) This mechanism of forming a cool phase in a hot atmosphere cannot be captured in cosmological simulations, because they are generally unable to resolve individual SN remnants.  

Future work is required to better understand this mechanism within more accurate conditions featuring gravitational stratification, magnetic fields, conduction and other relevant physical phenomena.

\vspace{0.2in}

\section*{Acknowledgement}
We thank the referee for the helpful comments which improved the clarity of the paper. We thank members of the SMAUG collaboration for useful discussions. ML thanks the helpful discussion with Feng Yuan and Mark Voit. Computations were performed using the publicly-available Enzo code, which is the product of a collaborative effort of many independent scientists from numerous institutions around the world. Their commitment to open science has helped make this work possible. Data analysis and visualization are partly done using the \textsf{yt} project \citep{turk11}. The simulations are performed on the Rusty cluster of the Simons Foundation and the XSEDE clusters supported by NSF. We thank the Scientific Computing Core of the Simons Foundation for their technical support. We acknowledge financial support from NSF (grant AST-1615955, OAC-1835509 to GB, AST-1715070 to EQ), NASA (grant NNX15AB20G to GB), and the Simons Foundation (grant 510940 to ECO, 528306 and a Simons Investigator Award to EQ).

\appendix
\restartappendixnumbering

\section{Cooling curve}

\begin{figure}
\begin{center}
\includegraphics[width=0.5\textwidth]{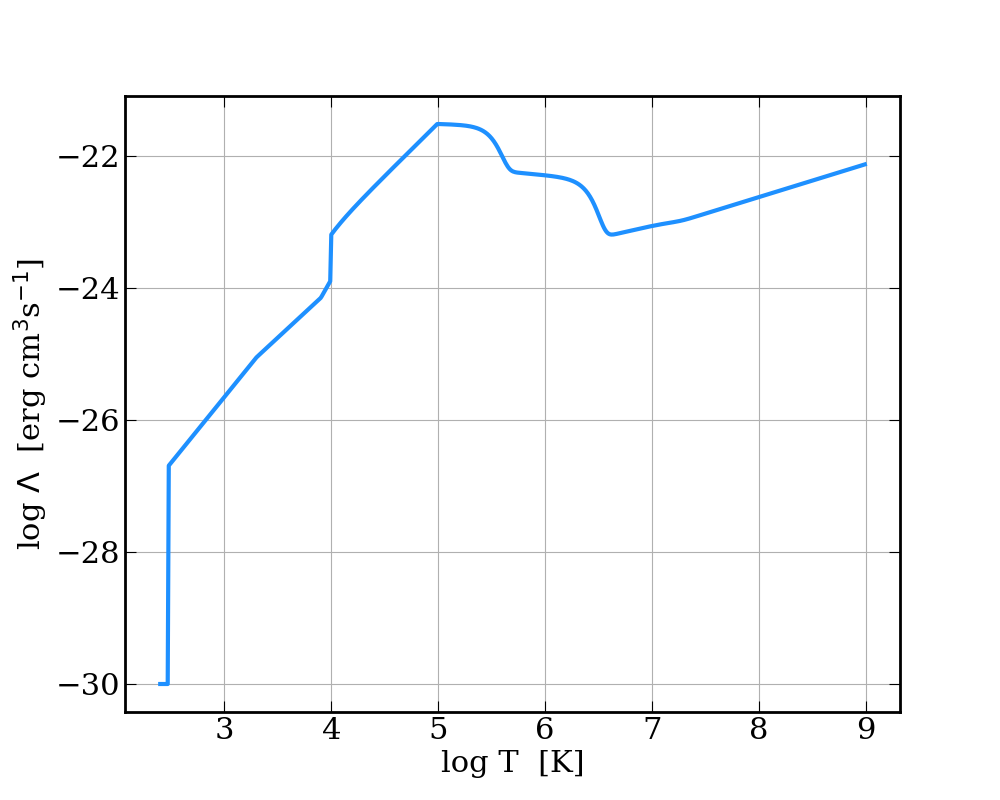}
\caption{Cooling curve adopted in this paper. }
\label{f:cooling_curve}
\end{center}
\end{figure}

Fig. \ref{f:cooling_curve} shows the cooling curve adopted in this paper. This is taken from \cite{rosen95}, with a metallicity of half the solar value.

\section{condition for forming a cooling shell of SNR} 
\label{sec:shell}
 
Below we discuss the critical condition for whether the cooling phase is present or not.
The evolution of a spherical blast wave in the energy-conserving phase is indicated by the Sedov-Taylor solution \citep{taylor50,sedov59},
\begin{equation}
R_{\rm{shock}}=\beta \left(\frac{E_{\rm{SN}}t^2}{\rho} \right)^{1/5},
\label{eq:rshock}
\end{equation}
where $R_{\rm{shock}}$ is the radius of the shock front, $E_{\rm{SN}}$ is the energy released by a SN, $\rho$ is the density of the ambient medium, and the constant $\beta \approx 1.15$ for an adiabatic index $\gamma=$5/3 \cite[e.g.][]{sedov59}. 

The cooling radius of a SN-induced blast wave can be obtained through the equation
\begin{equation}
t_{\rm{dyn}} =  t_{\rm{cool,s}}, 
\end{equation}
where $t_{\rm{dyn}}$ is the dynamical time of the shock from Eq. \ref{eq:rshock}, i.e., $(R^5\rho/E_{\rm{SN}}/\beta)^{0.5} $, and $t_{\rm{cool, s}}$ is the cooling time of the shock front, $3k_b T\mu/(4\rho\Lambda)$, in which $\mu$ is the mean molecular weight, $k_b$ is the Boltzmann constant, $\Lambda$ is the temperature-dependent cooling rate, and $T$ is the post-shock temperature. The term $4\rho$ is from assuming the strong-shock limit -- the density jump across the shock is a factor of 4. Assuming a power-law cooling curve $\Lambda(T)= 1.1\times 10^{-22} \rm{erg\ cm^3 s^{-1}} T_6^{-0.7}$ for $10^{5}< T < 10^{7.3}$K \citep{draine11}, where $T_6 = T/10^6$K, the shock radius at which cooling becomes important is
\begin{equation}
R_{\rm{cool,s}} = 23.7\rm{pc}\  n_1^{-0.42} E_{51}^{0.29},
\label{eq:rcool}
\end{equation}
where $n_1 = \rho/\mu/1 \rm{cm^{-3}}$, $E_{\rm{51}} = E_{\rm{SN}}/ 10^{51} \rm{erg} $.

The condition for a blast wave to fade away before it cools is simply 
\begin{equation}
R_{\rm{fade}} < R_{\rm{cool,s}}.
\label{eq:2r}
\end{equation}
Combining Eq. \ref{eq:rfade}, \ref{eq:rcool}, and \ref{eq:2r}, one gets
\begin{equation}
n < n_{\rm{crit}},
\label{eq: n}
\end{equation}
where
\begin{equation}
 n_{\rm{crit}}= 0.11\ \rm{cm^{-3}}\  (T_6/\alpha)^{3.85} E_{\rm{51}}^{-0.50}. 
 \label{eq:ncrit}
\end{equation}

Eq. \ref{eq:ncrit} shows $n_{\rm{crit}}$ increases sharply with $T$. For $T=10^4$ K, $n_{\rm{crit}}=2.2\times 10^{-9}$ cm$^{-3}$, whereas for $T=10^7$ K, $n_{\rm{crit}}=$ 780 cm$^{-3}$. So in the ISM of disk galaxies, the cooling shell forms first, whereas in the hot and tenuous ISM of elliptical galaxies, the the blast waves fade into sound waves first. In simulations presented in this paper, $T$ ranges from $3\times 10^6 - 3\times 10^7$ K, and $n$ from $2\times 10^{-3}$-0.3 cm$^{-3}$. We did not observe the cool shell formation at the shock fronts for any of our runs. This is consistent with Eq. \ref{eq: n}, \ref{eq:ncrit}. Note that the Sedov-Taylor solution we use here assumes that the internal energy of the ambient gas can be ignored. This is not quite true when the blast wave decays into a sound wave. For example, \cite{tang05} shows that the blast wave propagating in a hot medium deviates from the Sedov-Taylor solution at late stages. Nevertheless, it gives useful order-of-magnitude estimate of $n_{\rm crit}$.

\section{resolution check}
\restartappendixnumbering

\begin{figure}
\begin{center}
\includegraphics[width=0.5\textwidth]{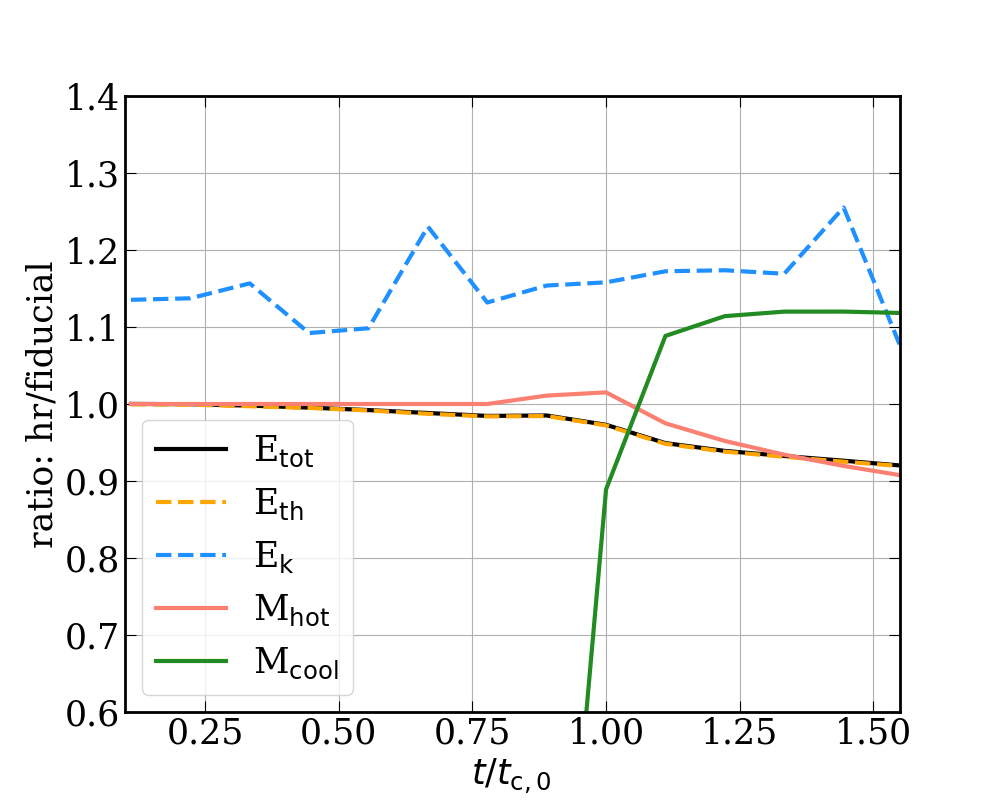}
\caption{Resolution check for n0.02-T3e6. The y-axis indicates the ratio of the compared quantities from the high-resolution run and those from the fiducial one.   }
\label{f:res_check}
\end{center}
\end{figure}

We run a high-resolution simulation n0.02-T3e6-hr, in which the spatial resolution is half the fiducial value. The high-resolution run lasts for 1.6 \tc\ (\tc$=$ 90 Myr). The formation time of the cool phase \tmulti\ for the high-resolution run, 82 Myr, agrees very well with that of the fiducial run, 79 Myr. Fig. \ref{f:res_check} compares several quantities, including the energy (total $E_{\rm{tot}}$, thermal $E_{\rm{th}}$ and kinetic $E_{\rm{k}}$) in the box, and the gas mass in hot ($T>2\times 10^4$ K) and cool ($T<2\times 10^4$ K) phases as a function of time. The thermal and total energy in the two runs start to deviate when it is close \tmulti , but overall they agree to within 10\%. The difference in kinetic energy starts early on, and remains at about 10-20\% level. The amount of cool gas also agrees at about 12\%, once it forms. The largest difference is the empirical mixing time, $t_d$ (listed in Table \ref{table1}): the high-resolution run has a longer $t_d \approx $ 28 Myr, compared to $t_d \approx $ 20 Myr for the fiducial case. This is due to the smaller numerical diffusion for the high-resolution case. Of course, the structure of the cool gas is under-resolved, even in the higher resolution simulations.

\vspace{0.3in}


\bibliographystyle{aasjournal}

\end{CJK*}

\end{document}